\documentclass[preprint]{aastex}

\slugcomment{}

\shorttitle{Nuclear Properties of Spiral Galaxies. }
\shortauthors{ M.~A.~Hughes,D. ~Axon, J.~Atkinson, A.~Alonso-Herrero, C. ~Scarlata et al.}

\begin{document}


\title{Nuclear Properties of Nearby Spiral Galaxies from Hubble Space Telescope NICMOS imaging and STIS Spectroscopy.\footnote{Based on observations with the NASA/ESA Hubble Space Telescope
obtained at the Space Telescope Science Institute, which is operated by the Association of Universities for Research in Astronomy, Inc., under NASA
contract NAS 5-26555.}}


\author{
M.~A.~Hughes\altaffilmark{1},
D.~Axon\altaffilmark{11},
J.~Atkinson\altaffilmark{1},
A.~Alonso-Herrero\altaffilmark{2},
C.~Scarlata\altaffilmark{7},
A.~Marconi\altaffilmark{4}, 
D.~Batcheldor\altaffilmark{1},
J.~Binney\altaffilmark{6}, 
A.~Capetti\altaffilmark{5},
C.~M.~Carollo\altaffilmark{7},
L.~Dressel\altaffilmark{3},
J.~Gerssen\altaffilmark{12},
D.~Macchetto\altaffilmark{3},
W.~Maciejewski\altaffilmark{4,10}, 
M.~Merrifield\altaffilmark{8},
M.~Ruiz\altaffilmark{1},
W.~Sparks\altaffilmark{3},
M.~Stiavelli\altaffilmark{3},
Z.~Tsvetanov\altaffilmark{9}, 
 }

\altaffiltext{1}{Centre for Astrophysical Research, STRI, University of Hertfordshire, Hatfield, Hertfordshire, AL10 9AB, UK.}
\altaffiltext{2}{Departamento de Astrofisica Molecular e Infrarroja, CSIC, Madrid, Spain.}
\altaffiltext{3}{Space Telescope Science Institute, 3700 San Martin Drive, Baltimore, MD 21218}
\altaffiltext{4}{INAF-Osservatorio Astrofisico di Arcetri, Largo E. Fermi 5, 50125 Firenze, Italy }
\altaffiltext{5}{INAF-Osservatorio Astronomico di Torino, I-10025 Pino Torinese, Italy.} 
\altaffiltext{6}{Oxford University, Theoretical Physics, Keble Road, Oxford, OX1 3NP, UK}
\altaffiltext{7}{Eidgenoessische Technische Hochshule Zuerich, Hoenggerberg HPF G4.3, CH-8092 Zuerich, Switzerland}
\altaffiltext{8}{School of Physics and Astronomy, University of Nottingham, NG7 2RD, UK}
\altaffiltext{9}{Center for Astrophysical Sciences, Johns Hopkins University, 239 Bloomberg Center for Physics
\& Astronomy, 3400 North Charles Street, Baltimore, MD 21218}
\altaffiltext{10}{Obserwatorium Astonomiczne Uniwersytetu Jagiello\'nskiego, Poland}
\altaffiltext{11}{Deptartment of Physics, RIT, 84 Lomb Memorial Dr., Rochester, NY 14623-5603}
\altaffiltext{12}{Department of Physics, University of Durham, Rochester Building,
Science Laboratories, South Road, Durham, DH1 3LE, UK}


\begin{abstract}

We investigate the central regions of 23 spiral galaxies using archival NICMOS imaging and STIS spectroscopy. The sample is
taken from our program to determine the masses of central massive black holes (MBH) in 54 nearby spiral galaxies. Stars are
likely to contribute significantly to any dynamical central mass concentration that we find in our MBH program and this paper
is part of a series to investigate the nuclear properties of these galaxies.  We use the Nuker law to fit surface brightness
profiles, derived from the NICMOS images, to look for nuclear star clusters and find possible {\it extended} sources in  3
of the 23  galaxies studied (13 per cent). The fact that this fraction is lower than that inferred from optical {\it Hubble
Space Telescope} studies is probably due to the greater spatial resolution of those studies.  Using R-H and J-H colors and
equivalent widths of H$\alpha$ emission (from the STIS  spectra) we  investigate the nature of the stellar population with
evolutionary models.   Under the assumption of hot stars ionizing the gas, as opposed to a weak AGN, we find that there are 
young stellar populations ($\sim$10--20\,Myr) however these data do not allow us to  determine what percentage of the total
nuclear stellar population they form. Also, in an attempt to find any  unknown AGN we use  [\ion{N}{2}]  and [\ion{S}{2}] 
line flux ratios (relative to  H{$\alpha$})  and find tentative evidence for weak AGN in NGC~1300 and NGC~4536.
\end{abstract}


\section{Introduction}

The most impressive result from studies of Massive Black Holes (MBHs) in recent years has been the discovery that the mass of the
central black hole is correlated with other properties of the galaxy, such as bulge mass (e.g. \citealt{kormendy1,marconi2}) and stellar velocity dispersion
\citep{ferrarese1,gebhardt1}. Consequently, it is prudent for any thorough investigation of MBHs to include a detailed study of the center of the
galaxy in question. With this in mind, this paper forms part of a series where we investigate the central regions of galaxies from
our Hubble Space Telescope (HST) program to find massive black holes in nearby spiral galaxies (GO:8228, PI: D.~Axon). The sample
is limited to galaxies within a recessional velocity less than 2000~kms$^{-1}$ that are known to have H$\alpha$ emission from ground
based observations. The good spectra and images, and a description of the sample, are presented in \citet{hughes1} (Hereafter H03)
and \citet{scarlata1} (Hereafter S04). The first black hole mass estimates from the program are presented in \citet{marconi1}, for
NGC~4041 and Atkinson et al. (submitted), for NGC~1300 and NGC~2748. 

One aspect of galaxy centers, which may have important consequences for black hole mass measurements is the presence of nuclear
star clusters. HST observations have played a major role in showing that many (50\%+) spiral galaxies have identifiable cores which are 
photometrically and morphologically distinct from the surrounding bulge (e.g. \citealt{boeker2,carollo3}). Such compact luminous
sources are generally believed to be nuclear star clusters and, in some cases, this has been demonstrated with spectroscopy (e.g. \citealt{boeker1,walcher1,colina1}). The reason that these clusters may be important for black hole mass estimates is that invariably the region over which the central mass is determined includes the region occupied by the cluster. Thus, if the cluster mass is significant with respect to the black hole mass then it needs to be taken into account.

Although quantifying the mass of the central stellar populations is beyond the scope of this paper, we attempt to determine other useful properties of the central regions of the galaxies. Specifically, this paper is a companion to S04 in which we quantified the profile of the bulges from the STIS images using the NUKER law profile {\citep{lauer1} and looked for the presence of nuclear star clusters.  The first aim of  this paper is to re-use the NUKER profile and apply it to archival NICMOS images of galaxies from our sample. Since near-infrared images are not as affected by dust obscuration as the STIS images these provide the opportunity of producing a clearer view of the central bulge, and a better way of quantifying the shape of the bulge. The second purpose of this paper is to use the spectral information from our STIS program \citep{hughes1}  and the color information from both our STIS images and archival NICMOS images in an attempt to quantify the age of the central stellar population.

The structure of this paper is as follows, In \S\ref{data} we describe the selection of the sample of
spiral galaxies, the STIS   spectroscopic observations and the  reduction  of the archival NICMOS
images.  In \S\ref{prof} we produce surface brightness profiles from the NICMOS  images to look for
central sources. In \S\ref{pop} we investigate the ages of the stellar populations at the centers of 
galaxies using  both color information from the NICMOS  and STIS   images and STIS   spectroscopy.

\subsection{The Sample Selection } \label{data}

The complete list of galaxies is presented in   table~1 in H03 and in Table~\ref{little_sample}
(this paper) we list those galaxies for which we were able to obtain archival NICMOS  H-band images. 
All  the galaxies are  classified as late-type spirals and are nearby, having recessional velocities
of less than 2000\,km~s$^{-1}$.

The original purpose of the STIS images was to locate the exact position of the nucleus of each
galaxy so that spectroscopic apertures could be accurately placed. Such `acquisition images' were
taken using the F28X50LP longpass filter. They are optical (central wavelength 7230~\AA), and are
approximately equivalent to R-band. For each galaxy, 2 acquisition images were taken to acquire the
nucleus and  3 longslit apertures were placed  to determine the nuclear gas disk kinematics. The pixel
size is 0.05\arcsec pixel$^{-1}$, and the field of view is 5\arcsec$\times$5\arcsec. Integration times
of the images vary from 20 to 60 seconds. The STIS images were reduced by the Flight Software (FSW)
involving the subtraction of a single bias number and the removal of hot pixels. See chapter 8 of the
STIS Instrument Handbook. 

Spectra of the galaxy centers covered 5 emission lines [\ion{N}{2}]~(6549.9~\AA),
H$\alpha$~(6564.6~\AA), [\ion{N}{2}]~(6585.3~\AA) and [\ion{S}{2}]~(6718.3 \& 6732.7~\AA). These spectra
are being used to map the velocity fields of  nuclear gas disks so that  central mass concentrations can
be measured. The spectra are described in more detail in H03.  Briefly, for each galaxy three
slits were placed, one on the brightest central source (presumed to be the nucleus, unless obscured by
dust) with 2 other, parallel slits 0.1~arcsec either side. In this paper, we use these spectra for 
another purpose; to estimate the age of  stellar populations at the centers of the
galaxies~(see~\S\ref{pop}).

\begin{deluxetable}{lcccc} 
\tabletypesize{\footnotesize} 
\tablewidth{0pt}
\tablecaption{The sample of 23 spiral galaxies with archival NICMOS  images}
\tablehead{Galaxy & Morphological   &Nuclear	    & Rec. Vel.  & Inclination\\   
	Name  & Type 	   &  Activity      &  (km s$^{-1}$)    & (degrees)}

\startdata 
\objectname[]{NGC 0157} &  SAB(rs)bc    &   ...                       &1589 & 55 \\
\objectname[]{NGC 0289} &  SAB(rs)bc    &   ...                       &1451 & 40 \\
\objectname[]{NGC 1300} &  (R')SB(s)bc  &   ...                       &1409 &  49 \\
\objectname[]{NGC 2748} &  SAbc         &  (\ion{H}{2})               &1741 & 73 \\
\objectname[]{NGC 2903} &  SB(s)d       &  \ion{H}{2} (\ion{H}{2})    &626  & 56 \\
\objectname[]{NGC 2964} &  SAB(r)bc     &   \ion{H}{2} (\ion{H}{2})   &1446 &  58    \\
\objectname[]{NGC 3259} &  SAB(rs)bc    &   ...                       &1929 & 63 \\
\objectname[]{NGC 3310} &  SAB(r)bc pec & \ion{H}{2} (\ion{H}{2})     &1208 &  31\\
\objectname[]{NGC 3949} &  SA(s)bc      & \ion{H}{2} (\ion{H}{2})     &1021 &  56\\
\objectname[]{NGC 4030} &  SA(s)bc      &   ...                       &1475 & 40 \\
\objectname[]{NGC 4258} &  SAB(s)bc     & L/S1.9 (S1.9)               &674  &	72\\
\objectname[]{NGC 4303} &  SAB(rs)bc    & \ion{H}{2}/S2 (\ion{H}{2})  &1619 & 19 \\
\objectname[]{NGC 4389} &  SB(rs)bc pec &   ...                       &940  & 54 \\
\objectname[]{NGC 4527} &  SAB(s)bc     & \ion{H}{2}/L (T2)	      &1776 & 70 \\
\objectname[]{NGC 4536} &  SAB(rs)bc    & \ion{H}{2} (\ion{H}{2})     &1846 &  59\\
\objectname[]{NGC 5005} &  SAB(rs)bc    & S2/L (L1.9)	              &1153 &	67\\
\objectname[]{NGC 5054} &  SA(s)bc      &   ...                       &1704 & 54 \\
\objectname[]{NGC 5055} &  SA(rs)bc     & \ion{H}{2}/L (T2)	      &726  &	56 \\
\objectname[]{NGC 5248} &  (R) SB(rs)bc & S2/\ion{H}{2} (\ion{H}{2})  &1248 & 51  \\
\objectname[]{NGC 5879} &  SA(rs)bc?    & L (T2/L2)	              &1049 &	74\\
\objectname[]{NGC 6384} &  SAB(r)bc     & L (T2) 	              &1780 &60	\\
\objectname[]{NGC 6951} &  SAB(rs)bc    & L/S2 (S2)	              &1705 &42	\\
\objectname[]{NGC 7331} &  SA(s)b       & L (T2)	              &992  &	75  \\
\enddata
\tablecomments{Morphological classification and activity status taken from NASA Extragalactic Database (NED) with classification shown in
parenthesis taken from \citet{ho1}, where \ion{H}{2}=\ion{H}{2} nuclei, S=Seyfert nuclei, L=LINER, T=Transition nuclei. `Rec. Vel.' is the recessional velocity corrected for Virgo-centric infall (LEDA). Inclination is the angle between the polar axis and the line of sight, in degrees (LEDA).}
\label{little_sample}
\end{deluxetable}    
\clearpage

The {\it HST} archive was used to search for NICMOS  observations  for the galaxies in our sample. We
found NIC2 F160W images (1.60\micron, similar to a ground-based H-band filter) for 23 galaxies, and for 
6 galaxies we found NIC2  F110W filter  images (1.10\micron, similar to a ground-based J-band filter). 
Many of the images were observed as part of an {\it HST} snapshot programme by members of our team and
appear in \citet{carollo3}. The pixel size of the images is 0.076''\,pixel$^{-1}$, which  produces a
field of view of 19.2\arcsec$\times$19.2\arcsec. The images were taken as part of a number of programs,
with  integration times of between 384 and 640 seconds. The images were reduced using NicRed
(\citealt{mcleod1}).  The  data reduction involved subtraction of the first  readout, dark current
subtraction on a readout-by-readout basis, correction for linearity and cosmic ray rejection (using {\sc
fullfit}), and flat fielding.  In-orbit darks with sample sequences and exposure times corresponding to
those of the observations were obtained from  other programs close in time. Generally between 10 to 20
darks were averaged together (after  subtraction  of the first readout) for a each sample sequence. If
more than one exposure for a given galaxy was taken,  the images were  shifted to a common position
using fractional pixel offsets and combined to create the final images. Since the images were well sampled, any loss in resolution through using this technique should not be significant.

\section{Surface Brightness Profiles and the Presence of Central Sources} \label{prof}

Surface brightness profiles have been used as an important diagnostic for inferring the presence of
resolved, central sources in galactic nuclei. \citet{boeker2} presented SBPs  derived from {\it WFPC2}
F814W images  and found many well resolved central sources. The presence of a central source is
indicated by an inflexion in the profile at small radii, see figure~\ref{ideal_profile}. Without
spectroscopy, such as performed by \citet{boeker1} on NGC~4449 (who found a 6-10\,Myr stellar
population) or \citet{walcher1}, we cannot be confident of the nature of these central sources. However,
the general assumption by most authors is that they are probably star clusters.

\begin{figure*}
\epsscale{1}
\plotone{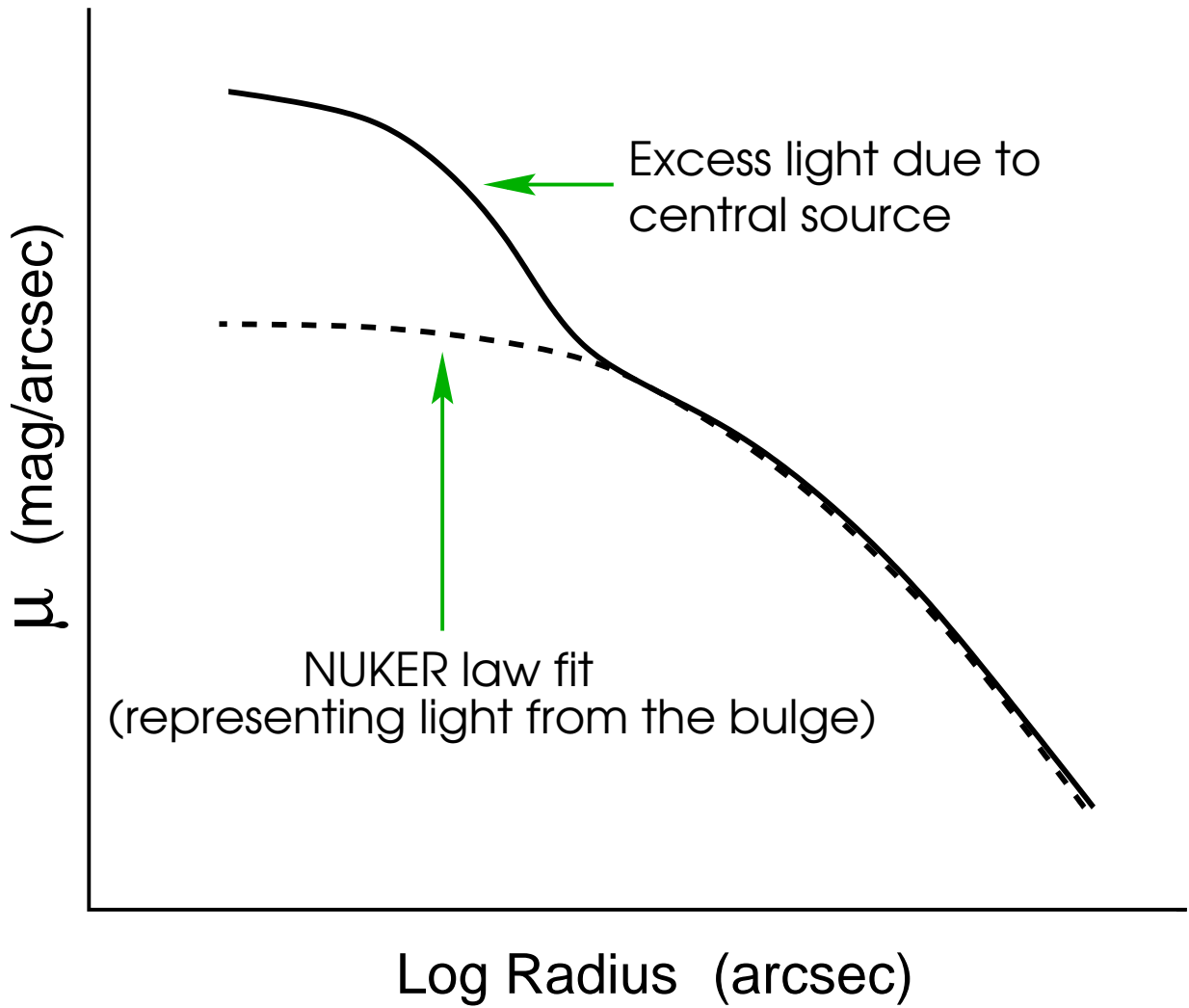}
\caption{An idealised picture of a surface brightness profile (SBP). In several cases, a clear bump is seen that cannot be accounted for in the standard Nuker law model.
This excess light is assumed to be from a compact luminous source, possibly a nuclear star cluster. }
\label{ideal_profile}
\end{figure*}

Over the past few years, analysis of central sources has been performed for a large number of disk
galaxies using {\it HST} images, in particular using {\it WF/PC~1} (\citealt{phillips1}), {\it WFPC~2}
(\citealt{carollo2,carollo1}) and NICMOS  (\citealt{carollo3,seigar1}). The main inference of these studies has
been that central sources are a common feature of galaxy centers.

\begin{figure*} 
\epsscale{0.85}
\plotone{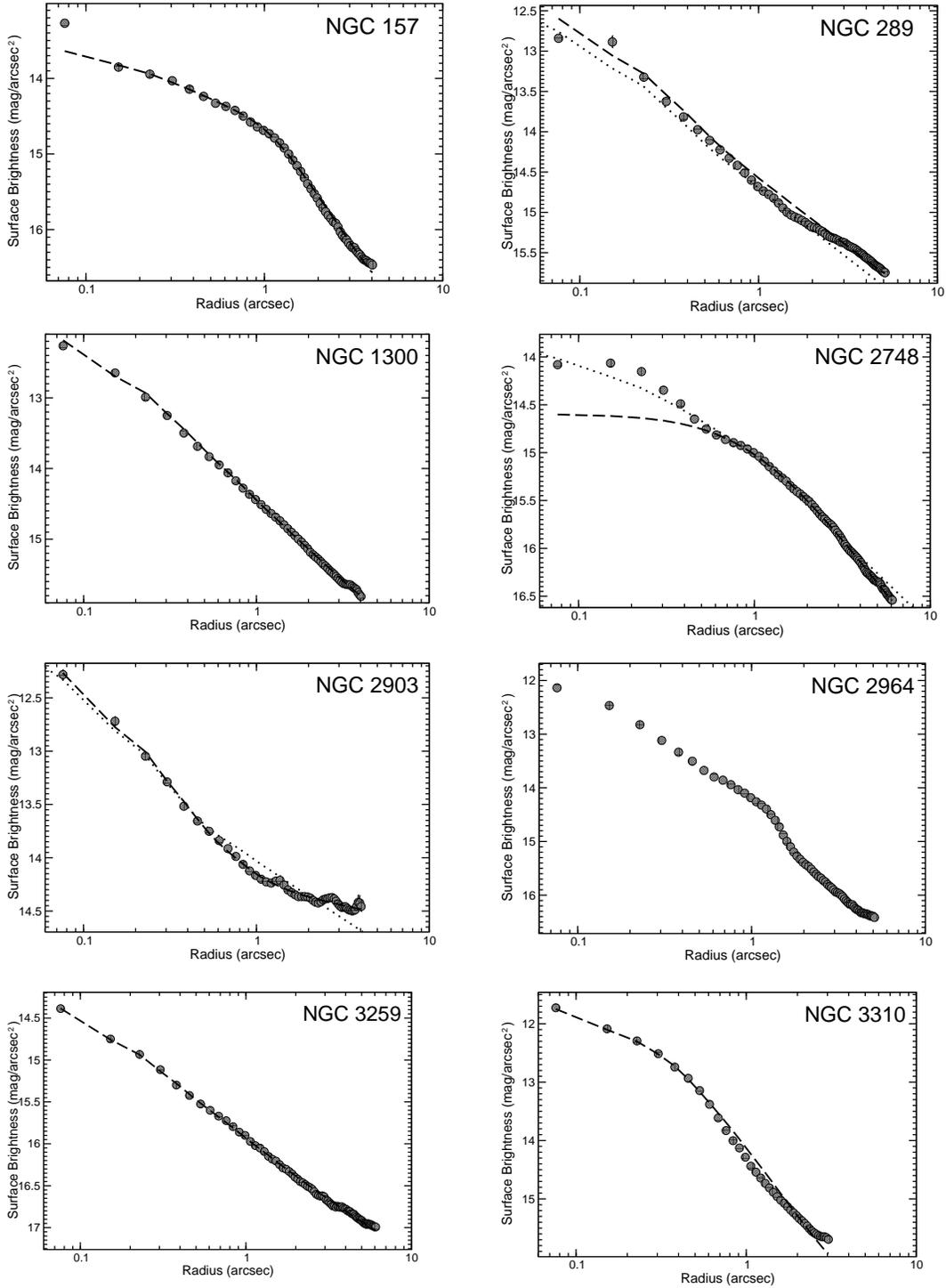} 
\caption{NICMOS surface brightness profiles for NGC~157, NGC~289, NGC~1300, NGC~2748, NGC~2903, NGC~2964, NGC~3259 and NGC~3310. Filled circles are the surface brightness profiles found from fitting ellipses to the isophotes. Dashed lines are our Nuker law fits to the profile and the dotted lines (NGC~289, NGC~2748 and NGC~2903)  are the Nuker profile fits of \citet{seigar1},which which have been convolved with our {\sc Tiny Tim} \citep{krist1} generated PSFs.}
\label{profiles1}
\end{figure*}

\subsection{Nuker Surface Brightness Profile Fits}

We used the {\sc ellipse} program under {\sc iraf}\footnote{IRAF is distributed by the National Optical Astronomy
Observatories, which are operated by the Association of Universities for Research in Astronomy, Inc., under
cooperative agreement with the National Science Foundation.}  to perform isophotal ellipse fitting on the NICMOS images,
from which we produced surface brightness profiles. In each case we  allowed position angle and ellipticity to vary but kept the position of the center fixed.

The flux calibration  of the NICMOS images was performed using conversion factors  (from ADU\,s$^{-1}$ to Jy) based on
measurements of the standard star P330-E, taken during the Servicing Mission  Observatory Verification  program (M. J.
Rieke 1999, private communication). Consequently, results from the ellipse fitting were converted to magnitude per
arcsec$^{2}$ by:

\begin{equation}
m_{H} = -2.5\log\left ( \frac{2.19\times10^{-6}}{1083} \ counts \ s^{-1} \right ) +5\log(scale)
\label{mag_eq}
\end{equation}

Where {\it scale} is the plate scale and is 0.076 \arcsec pixel$^{-1}$. For our J-band calculations (see table~\ref{allsample}) we substituted the factor  $ -2.5\log\left ( \frac{2.031\times10^{-6}}{1775} \ counts \ s^{-1} \right )$ into 
the first part of equation~\ref{mag_eq}. The uncertainty on the photometry is assumed to be 2\%.

 To minimise assumptions, we  chose to {\it not}   mask out complicated structures such as star formation rings.
Surface brightness profiles from  the NICMOS  images of NGC~289, NGC~2748, NGC~2903, NGC~3259, NGC~3949, NGC~4527,
NGC~4536 and NGC~6384 have previously been produced, from the same raw NICMOS  images, by \citet{seigar1}. Since these
galaxies are also in our sample, we re-do the profile fitting here. This enables us to perform a check on the
consistency between both results. Following \citet{seigar1} the SBPs were fitted with a Nuker law profile, of the form introduced by \citet{lauer1}:

\begin{equation}
I(r) = 2^{(\beta-\gamma)/\alpha}I_{b} \left(\frac{r_{b}}{r}\right)^{\gamma}  \left[ 1+ \left( \frac{r}{r_{b}} \right)^{\alpha}
\right]^{(\gamma-\beta)/\alpha}
\label{nuker}
\end{equation}

where, $\beta$ indicates the steepness of the outer profile, $\gamma$ measures the steepness of the inner profile,
$\alpha$ indicates the sharpness of the transition between these two profiles, while $r_{b}$ is the point of this
transition at which the SBP has a brightness of $I_{b}$.

\begin{figure*}
\epsscale{0.75}
\plotone{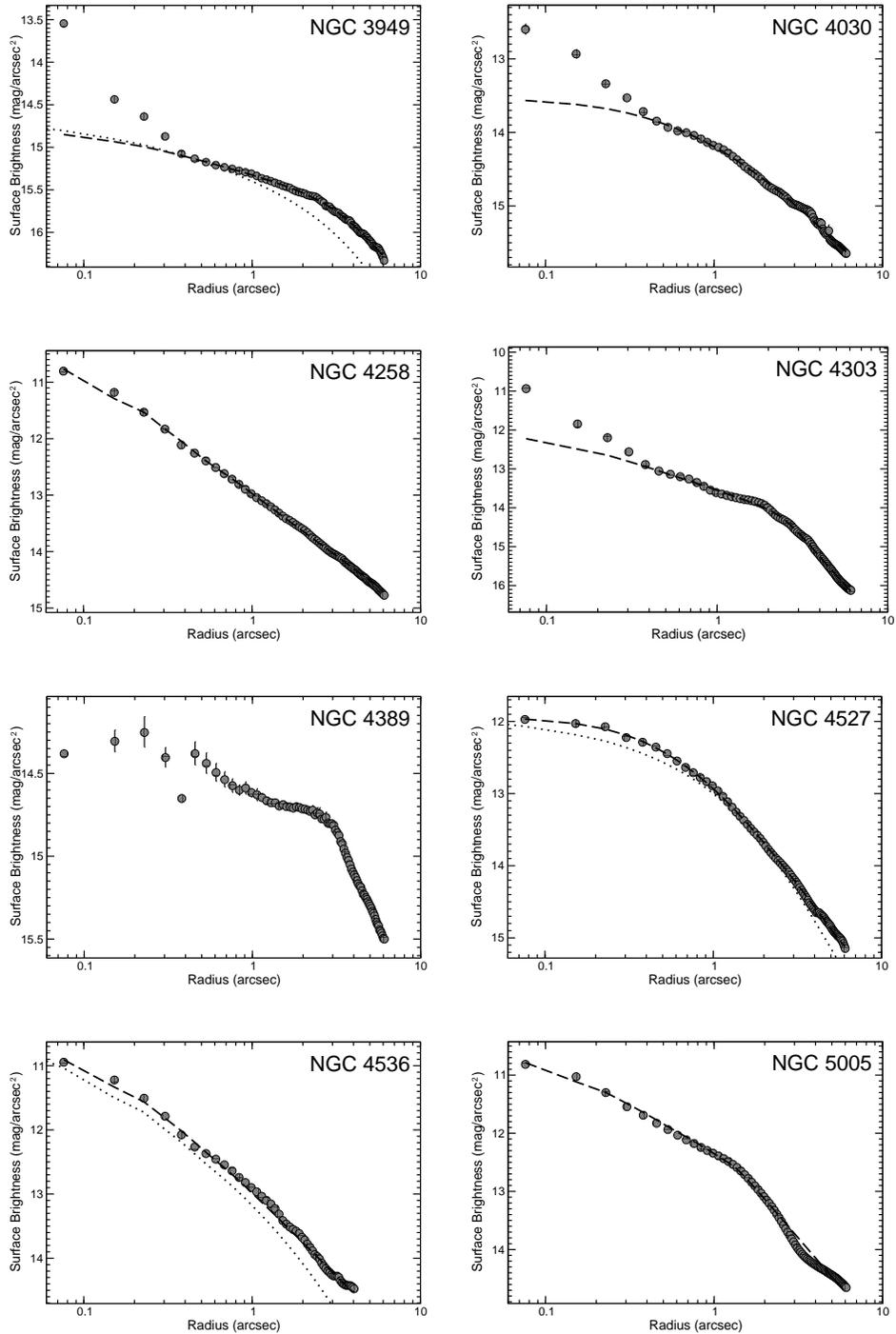} 
\caption{NICMOS  surface brightness profiles for NGC~3949, NGC~4030, NGC~4258, NGC~4303, NGC~4389, NGC~4527, NGC~4536 and NGC~5005. The profiles have been fit with a Nuker law. Dashed lines are our fits and dotted lines (NGC~4527 and NGC~4536) are those of \citet{seigar1}, which which have been convolved with our {\sc Tiny Tim} \citep{krist1} generated PSFs. Variations between the dotted lines and the surface brightness profiles are probably due to differing assumptions made during the isophotal ellipse fitting of the NICMOS images. }
\label{profiles2}
\end{figure*}

It is important to take proper account of the NICMOS point spread function when fitting the surface brightness profiles. This can be done by either deconvolving the images prior to the fit or by convolving each model profile with a suitable psf before the fit is made. Unfortunately, deconvolution tends to increase the resultant noise. For this reason, we chose to follow the latter method, which has also been favored by other authors recently (e.g. \citealt{seigar1,scarlata1}). Consequently, the models were  convolved with  {\sc Tiny Tim}  generated NICMOS point spread functions  \citep{krist1} before attempts were made to fit the surface brightness profiles.

The fitting routine is fundamentally the same as the one we used in \citet{scarlata1}. Briefly, we constructed an IDL program that made use of  the standard IDL procedure {\it curvefit}, which minimses  $\chi^{2}$ by using the Levenberg-Marquardt method to search for the best fit. The user is able to supply the inner radius at which the fit is started as well as the  PSFs that are convolved with the models before each fit is attempted.  As in S04, rather than add an additional model to account for central sources, we varied the inner radial range until we arrived at a suitable fit that excluded the central region and only fit the bulge. In some cases (NGC~157 and NGC~5879) we could not clearly identify a {\it resolved} central source, but still found that the fits were improved by varying the inner radius. Possible resolved central sources are treated separately in section~\ref{quantify}.

The best profile fits are shown in Figs.~\ref{profiles1}, \ref{profiles2} and
\ref{profiles3}. Where possible both the \citet{seigar1} fits (dotted line --converted to Johnson magnitudes) and our fits
(dashed line) are shown together. Following \citet{seigar1} we corrected for extinction using the data of
\citet{burstein1}. 

 Table~\ref{profile_fits} lists the parameters of the fits.  For NGC~2964 and NGC~4389 we did not find a good fit to the
surface brightness profile. In many cases, however, the SBP can be completely  described by the Nuker law. Parameters from
Nuker law fits performed by \citet{seigar1} are also included in  the table.  Usually we found the value of  $\alpha$ to be
poorly constrained. This is unsurprising since, in general, the range  of points in the profile fits extend well beyond the
break radius, $r_{b}$.  This is important because as  $r>>r_{b}$, the  factor, $\left[ 1+ \left( \frac{r}{r_{b}}
\right)^{\alpha} \right]^{(\gamma-\beta)/\alpha}$ tends to become independent of $\alpha$.

 The main reason for the differences in this work and that of \cite{seigar1} is the shape of the
surface brightness profiles. While in part this is due to the decision as to whether or not the
profiles should be smoothed, and which regions to mask, it may also be  due to the subjective
choices (e.g. fixing ellipticity) that have to be made when performing the isophote fits. An
example of the necessary subjectivity in surface brightness profile fitting can be seen for
NGC~2748. We originally interpreted the profile as showing evidence for a nuclear cluster and so start the
fit at an inner radius of 0.68\arcsec. In contrast, \citet{seigar1} start the profile fit inward
of this and so it is unsurprising that the Nuker parameters should be different. Generally, a
decision has to be made for each profile as to the region where the Nuker profile is valid. This
should be inbetween any central source and the region where the galactic disk dominates. We are
assuming that the Nuker profile, which \citet{lauer1} originally proposed to describe elliptical
galaxies is actually a good description of the bulges in spiral galaxies. In some cases it is an
over-complicated description of the bulge profile, while in other cases structure in the profile
can produce a wide range of parameters,  depending on the subjective choice as to the region over
which the fitting is performed.  It is also important to note  that dust could give the misleading
impression that a cluster is present. For example if we observe a SBP such as shown in
figure~\ref{ideal_profile} we must be careful that the inflexion between the apparent bulge and
central source profile is not in reality  a result of dust obscuration.

\begin{figure*}
\epsscale{0.85}
\plotone{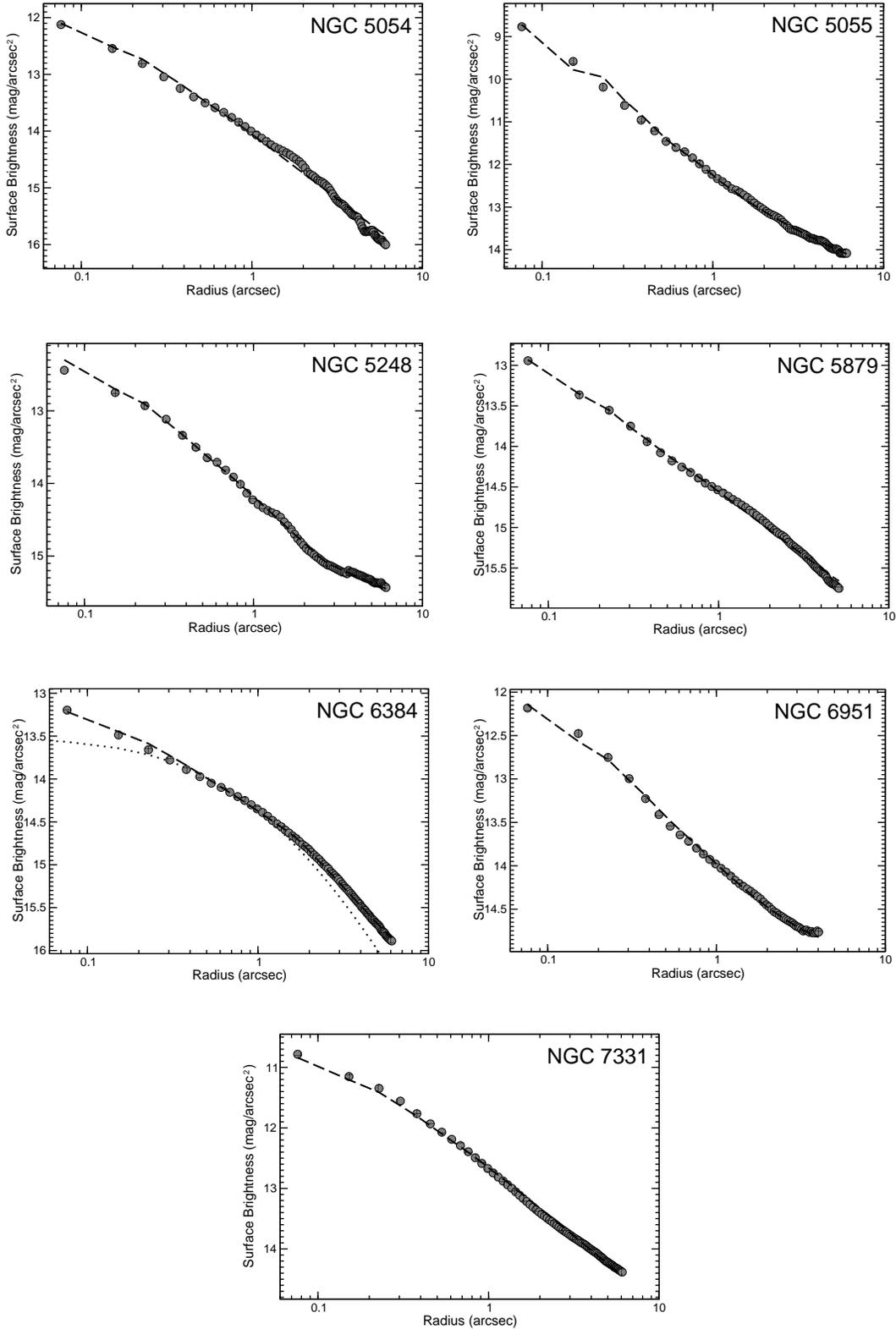} 
\caption{NICMOS  surface brightness profiles for NGC~5054, NGC~5055, NGC~5248, NGC~5879, NGC~6384, NGC~6951 and NGC~7331. Dashed lines are our fits and dotted lines (NGC6384) are those of \citet{seigar1}, which have been convolved with our {\sc Tiny Tim} \citep{krist1}  generated PSFs. The kink in the profile of NGC~5055 at small radii is due to the convolution of the NICMOS psf with a bright point source.}
\label{profiles3}
\end{figure*}

\begin{figure*}
\epsscale{0.45}
\plotone{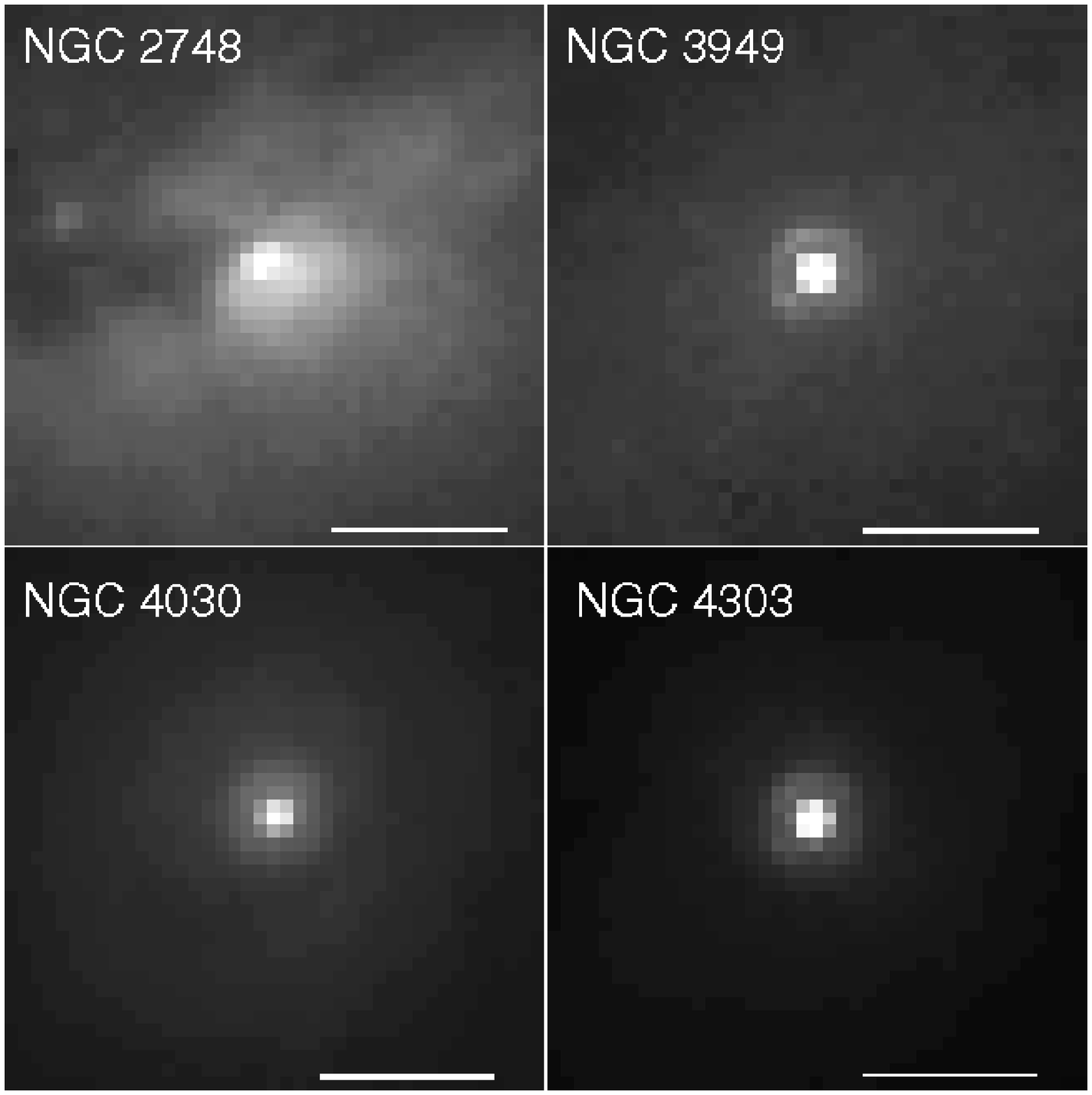} 
\caption{ The archival NICMOS f160w images for the 4 central source candidates. Cut levels have been adjusted to highlight the nuclei. The white bar shows a scale of 1~\arcsec. The image for NGC~2748 shows that the presence of significant amounts of dust is the likely cause of the inflexion in the surface brightness profile. The central sources in the other galaxies are not much different to the size of the NICMOS PSF (see also figure~\ref{psf}), as can be seen by the presence of Airy rings}
\label{images}
\end{figure*}

It is useful to compare which central sources we detect, or fail to detect,  with the results we presented in S04. For 4 of
the galaxies; NGC~4536, NGC~5054, NGC~6384 and NGC~6951, we found central sources in the STIS   profiles (shown in table 3
of S04), yet adequate Nuker law fits are made in figures~\ref{profiles2} and ~\ref{profiles3} without the need to invoke the
presence of a star cluster.  There are several possibilities to explain why this situation might arise: 1) The population
could be blue and hence not readily visible in the NICMOS  images. 2) The presence of dust in the STIS   images may have
given us the misleading impression of a cluster being present 3) An older, redder population may dominate the NICMOS  images
to an extent that the cluster is not visible 4) The lower resolution of the NICMOS  images may have smoothed out any image
of the cluster.

\begin{deluxetable}{lcccccccccccc} 
\tabletypesize{\scriptsize} 
\tablewidth{0pt}
\tablecaption{Parameters from NUKER law fits.}
\tablehead{NGC &     range     &   & $\alpha$	  &  &  $\beta$       &       &  $\gamma$    &   &  R$_{b}$	   &   & I$_{b}$	     &  \\      
               & ($\arcsec$)   &   &             &  &                &       &              &   &   ($\arcsec$)   &   & (mag/$\arcsec^{2}$) & 	   }

\startdata 

0157 & {0.15-6.08}   & {---}		& {6.83}    & {---}	   & {1.38}      & {---}     & {0.45} & {---}	   & {1.05}  &  {---}       & {14.69}	& {---}   \\
0289 & {0.08-6.08}   & {(0.0-5.0)}      & {47.67}   & {(12.22)}    & {0.67}      & {(0.69)}  & {0.92} & {(0.91)}   & {0.72}  &  {(0.60)}    & {14.34}	& {(15.68)} \\
1300 & {0.08-6.08}   & {---}		& {73.20}   & {---}	   & {0.89}      & { ---}    & {0.99} & {---}      & {0.95}  &  {---}       & {14.42}	& {---}       \\
2748 & {0.68-6.08}   & {(0.0-8.0)}      & {1.75}    & {(5.01)}     & {0.97}	 & {(0.71)}  & {0.00} & {(0.46)}   & {0.89}  &  {(0.86)}    & {14.94}	& {(16.31)} \\
2903 & {0.08-4.03}   & {(0.0-5.0)}      & {2.92}    & {(20.22)}    & {0.13}      & {(0.43)}  & {0.99} & {(0.95)}   & {0.74}  &  {(0.43)}    & {14.02}	& {(14.99)} \\
3259 & {0.08-6.08}   & {(0.3-8.0)}      & {0.23}    & {(7.33)}     & {0.14}      & {(0.68)}  & {1.29} & {(0.44)}   & {0.23}  &  {(0.00)}    & {14.86}	& {(15.60)} \\
3310 & {0.08-2.51}   & {---}		& {88.74}   & {---}	   & {1.51}      & {---}     & {0.74} & {---}	   & {0.38}  &  {---}       & {12.60}	& {---}       \\
3949 & {0.46-6.08}   & {(0.3-8.0)}      & {3.07}    & {(1.53)}     & {1.10}      & {(2.57)}  & {0.23} & {(0.26)}   & {3.74}  &  {(7.44)}    & {15.86}	& {(18.39)} \\
4030 & {0.61-6.08}   & {---}		& {0.88}    & {---}	   & {1.12}      & {---}     & {0.00} & {---}	   & {1.12}  &  {---}	    & {14.25}	& {---} 	\\
4258 & {0.08-6.08}   & {---}		& {32.94}   & {---}	   & {0.87}      & {---}     & {1.01} & {---}	   & {0.48}  &  {---}	    & {12.28}	& {---} 	\\
4303 & {0.53-6.08}   & {---}		& {10.13}   & {---}	   & {2.00}      & {---}     & {0.62} & {---}	   & {2.38}  &  {---}	    & {14.23}	& {---} 	\\
4527 & {0.08-6.08}   & {(0.0-8.0)}      & {1.33 }   & {(1.61)}     & {1.33}      & {(1.79)}  & {0.01} & {(0.27)}   & {0.58}  &  {(1.51)}    & {12.46}	& {(14.76)} \\
4536 & {0.08-4.03}   & {(0.0-8.0)}      & {0.39}    & {(2.13)}     & {1.26}      & {(2.06)}  & {0.67} & {(0.91)}   & {0.59}  &  {(2.43)}    & {12.44}	& {(15.76)} \\
5005 & {0.08-3.04}   & {---}		& {130.83}  & {---}	   & {1.34}      & {---}     & {0.71} & {---}	   & {1.37}  &  {---}	    & {12.59}	& {---} 	\\
5054 & {0.08-6.08}   & {---}		& {0.04}    & {---}	   & {1.62}      & {---}     & {0.14} & {---}	   & {0.39}  &  {---}	    & {13.17}	& {---} 	\\
5055 & {0.08-6.08}   & {---}		& {1.02}    & {---}	   & {0.58}      & {---}     & {1.62} & {---}	   & {1.08}  &  {---}       & {12.41}	& {---} 	\\
5248 & {0.08-3.04}   & {---}		& {98.30}   & {---}	   & {0.41}      & {---}     & {0.85} & {---}	   & {2.66}  &  {---}	    & {15.09}	& {---} 	\\
5879 & {0.15-5.09}   & {---}		& {8.16}    & {---}	   & {0.63}      & {---}     & {0.87} & {---}	   & {0.27}  &  {---}	    & {13.63}	& {---} 	\\
6384 & {0.08-6.08}   & {(0.3-8.0)}      & {452.89}  & {(0.76)}     & {0.85}      & {(1.78)}  & {0.54} & {(0.01)}   & {1.7 }  &  {(2.00)}    & {14.74}	& {(16.32)} \\
6951 & {0.08-4.03}   & {---}		& {1.33}    & {---}	   & {0.00}      & {---}     & {0.88} & {---}	   & {3.11}  &  {---}	    & {14.79}	& {---} 	\\
7331 & {0.08-6.08}   & {---}		& {0.11}    & {---}	   & {1.62}      & {---}     & {0.19} & {---}	   & {1.94}  &  {---}	    & {13.22}	& {---} 	\\

\enddata
\tablecomments{The Nuker law parameters as defined in Eq.~{\ref{nuker}}. Range is the region, in arcseconds, over which the fits were made. I$_{b}$  has been corrected using \citet{burstein1} results. The values in  parenthesis are taken from \citet{seigar1} and are presented here for comparison. Note: the quoted  $I_{b}$  values from \citet{seigar1} remain in the AB magnitude system, while ours are quoted in the Johnson system.}
\label{profile_fits}
\end{deluxetable}

The presence of central sources seems clear for  NGC~2748, NGC~3949, NGC~4030 and NGC~4303, where a  well sampled\footnote{The
surface brightness profile for NGC~157 also shows an excess above the Nuker fit, but in this analysis the excess is defined by a single point
only. Since our goal is to find resolved sources we did not investigate it further.} upturn in the profile  in the NICMOS  images is present.
However, closer inspection of the NICMOS images in figure~\ref{images} shows that dust is the likely cause of the inflexion in the SBP
for NGC~2748. This highlights the problem of using just the surface brightness profile to find central sources. Central sources for all of the remaining galaxies were previously reported in S04, though only the cluster in NGC~4030 was identified as resolved in
that paper\footnote{Note that in S04 point sources were defined to be those that have FWHM $\leq$ (1.5 $\times$ FWHM$_{PS}$), where FWHM$_{PS}$ means for a point source. Here we consider all sources with FWHM larger than the PSF to be at least, marginally resolved. Consequently, if we applied the S04 definition NGC3949 and NGC4303 would also be classified as point sources in Table~{\ref{quantify_source}}}. A central star cluster has also been identified by \citet{colina1} for NGC~4303, who find a 3.1~pc cluster of 4~Myrs in age. In
figure~\ref{psf} the SBPs are compared with the NICMOS  point spread function, and appear to be resolved. In these cases, if the Nuker law
describes the underlying galactic light, the magnitude of the central source can be extracted.

\subsection{Quantifying the Size and Luminosity of the Central Sources} \label{quantify} We attempted to quantify the size and luminosity of the central
sources.  Specifically, the techniques we used are similar to those described in S04, which were derived from the analysis of \citet{carollo1}. The
results are shown in Table~{\ref{quantify_source}}. In summary, the luminosity of the central source is bounded by the results from two methods. In the
first method a Gaussian profile is fit to each central source using the {\sc IRAF} task {\sc n2gaussfit}\footnote{The sides of the box, over which the Gaussians were fit to the images, were set to $\sim$ 8$\times$ FWHM of the NICMOS PSF.}. This task also calculates the background and so we produced background-subtracted model Gaussian central sources, from which the luminosity could be determined. In the second method,  the profile fits from
Table~\ref{profile_fits}  were used to produce model images of the galaxy bulges. These were then subtracted from the original galaxy images and the
luminosity of the central source was determined by using an aperture size of approximately 2$\times$FWHM  of the central
source.

As in S04 we used the Plummer law to determine the half-light radius, $b$, of the central sources.

\begin{equation}
I_{P}(r)=\frac{L}{\pi b^{2}}\left(1+\frac{r^{2}}{b^{2}}\right)^{-2}
\label{plummer}
\end{equation} 

Here, $I$ is the total luminosity of the central source and $r$ is the radius from the center of the source.  We do not directly measure $b$, instead we determine the FWHM from Gaussian fits to the central sources. To convert our results to the half-light radius $b$, we repeat the method described in S04. Briefly, we  established the relationship between the Gaussian FWHM and $b$ by  simulating multiple central sources of known $b$. These artifical images of star clusters were convolved with a NICMOS PSF and the FWHMs determined in the same way as the actual images. The uncertainties on $b$ were found by varying the sides of the box, over which the Gaussians  were fit to the original images,  by $\pm$~2~pixel

\begin{deluxetable}{ccccccc} 
\tabletypesize{\scriptsize} 
\tablewidth{0pt}
\tablecaption{Properties of the Central Sources.}
\tablehead{    Galaxy    &   M$_{H}$  & M$_{R}$     & b 	  & b$_{R}$     & FWHM   & FWHM (R)   \\      
                         &    (mag)   &  (mag)      & (pc)  &  (pc)       &  (arcsec)      &  (arcsec)   \\
                (1)      &     (2)    &  (3)        &  (4)  &  (5)        &  (6)   &  (7)        }

\startdata 

NGC 3949        &   -14.1$\pm$0.2   & ...		   & $1.7^{+0.2}_{-0.4}$       & PS        		&  0.17   &  ...     \\
NGC 4030 	&   -16.6$\pm$0.2   & -14.0 $\pm$ 0.3       &  $9.8^{+0.9}_{-1.1}$    & 7.6 $\pm$ 0.2 	&  0.32    &  0.16    \\
NGC 4303 	&   -18.0$\pm$0.2   & ...			   &  $4.5^{+0.2}_{-0.2}$      & PS             	 &  0.19   &  ...     \\
	
\enddata
\tablecomments{(1): The name of the galaxy. (2): The absolute H-band magnitude. (3): The equivalent absolute STIS R-band magnitude from S04. (4) The half-light radius from the plummer law shown in equation~\ref{plummer}. The value in parenthesis is from the STIS R-band results of S04, converted to the H$_{0}$ used here. PS means point source. (5): The half-light radius from S04.  (6): The Full Width Half-Maximum from the Gaussian fits. (7): The Full Width Half-Maximum from S04. Note that values from S04 have been converted using H$_{0}$=75 km s$^{-1}$}
\label{quantify_source}
\end{deluxetable}

\section{The Age of the Central Stellar Populations} \label{pop}

The main goal of our STIS   project is to understand the nature of the central regions  of a sample of nearby late type spirals, and  to determine
whether  the galaxies host a black hole in their centers. Using the STIS  spectroscopic data we are carrying out  detailed modelling of the
central velocity field of the nuclear disks in our sample of galaxies to infer the central mass concentrations. Before we  can conclude that the
central mass is a black hole we  need to determine  whether such a central mass concentration could be accounted for by a stellar population. 

As a first step towards finding the masses of the central stellar populations we need to quantify their ages. Due to the difficulties in
identifying nuclear clusters from the NICMOS images, and since some star clusters may be present but completely hidden by the bulge light we
decided to examine the central stellar population within two arbitrary, but small, apertures of diameters 50\,pc and 100\,pc. Galaxies for which
no obvious central source was seen in the NICMOS SBP  are also included. The magnitudes within these circular apertures, plus magnitudes within
annuli of the same area directly surrounding the apertures are shown in Table~{\ref{allsample}.

The near infrared  data are useful for investigating the masses of stellar populations for two main reasons. First,  the extinction in the
$H$-band is   greatly reduced in comparison with that in the optical ($A_H = 0.175 A_V$, \citealt{rieke1}). Second,  the infrared mass-to-light
ratios are less dependent  on the galaxy properties than the optical ones. Still they can vary by a factor of  2 (see e.g., \citealt{bell1}), so
an estimate of the age of the stellar population  is necessary.

 We used Starburst99 (\citealt{leitherer1})  evolutionary synthesis models. Our goal is not to constrain the IMF, so we  used just the standard
Salpeter IMF with an upper mass cutoff of  $M_{\rm up} = 100\,{\rm M}_\odot$. The model assumes solar metallicity and  instantaneous  star
formation. For the 6 galaxies with two observed colors (R-H and J-H) we can  plot a two color diagram and compare with the model outputs.  The
NICMOS  magnitudes are measured in the Johnson photometric system whereas the STIS  acquisition magnitudes are given  in the STMAG  photometric
system. We have used the Starburst99 spectral energy distribution outputs to convert the STIS  acquisition magnitudes to the R band using the
synphot  package in {\sc iraf}.

\begin{figure*}
\plotone{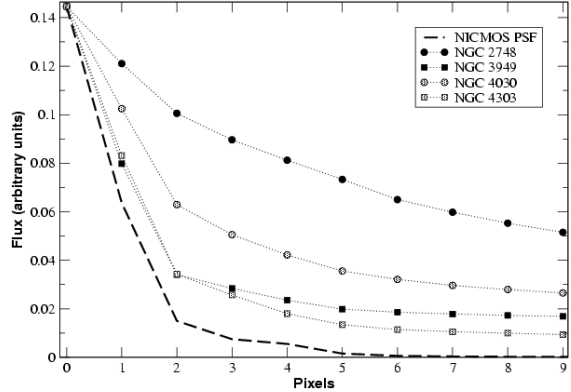}
\caption{Comparison of light profiles of those galaxies that seem to harbor compact sources with the NICMOS  point spread function (PSF). The NICMOS  profile was generated using the {\sc Tiny Tim} software package (\citealt{krist1}). The profiles have been scaled so that the fluxes are equal at the central pixel. No attempt has been made to extract the underlying bulge light.}
\label{psf}
\end{figure*}

\begin{deluxetable}{lcccccccccccc} 
\tabletypesize{\scriptsize} 
\tablewidth{0pt}
\tablecaption{Magnitudes within a circular aperture and equivalent widths of H$\alpha$}
\tablehead{{\bf Galaxy}&{\bf R}	&{\bf R}     &{\bf R(a)}&{\bf R(a)} &{\bf R-H }&{\bf R-H } &{\bf R-H(a)}&{\bf R-H(a)}&{\bf J-H} &{\bf J-H}  &{\bf EW(\AA)}  &{\bf EW(\AA)}\\ &{\bf 50pc} &{\bf 100pc} &{\bf 50pc}&{\bf 100pc}&{\bf 50pc}&{\bf 100pc}&{\bf 50pc}   &{\bf 100pc}  &{\bf 50pc}&{\bf 100pc}&{\bf 50pc}&{\bf 100pc}}

\startdata
NGC 0289     &18.40 & 17.38 & 18.98  & 17.93   & 3.61  & 3.61 & 3.64 & 3.36 & 1.14 & 1.17   & 20.1 (0.5)  & 18.6 (0.3)  \\
NGC 1300     &17.47 & 16.51 & 18.07  & 17.25   & 3.15  & 3.15 & 3.15 & 3.11 & ...  & ...   & 5.1 (0.1)   & 4.0 (0.1)	\\
NGC 2748     & ...  & ...   & ...    & ...     & ...   & ...  & ...  & ...  & ...  & ...   & 55 (2)	&  35.7 (0.9)	 \\
NGC 2903     &16.80 & 15.68 & 17.32  & 15.90   & 3.55  & 3.55 & 3.45 & 3.31 & ...  & ...   & 9.3 (0.1)   & 12.6 (0.2)  \\
NGC 2964     &17.61 & 16.57 & 18.14  & 17.16   & 3.31  & 3.31 & 3.24 & 3.18 & ...  & ...   & 10.6 (0.1)  & 16.7 (0.3)  \\
NGC 3259     &19.65 & 18.62 & 20.15  & 19.12   & 2.72  & 2.72 & 2.88 & 2.83 & ...  & ...   & 53.8 (0.7)  & 37 (1) 	 \\
NGC 3310     &16.42 & 15.47 & 16.92  & 16.59   & 2.92  & 2.92 & 2.86 & 2.84 & 0.93 & 0.92  & 75 (1)	 & 75 (1)	  \\
NGC 3949     &18.08 & 17.04 & 18.68  & 17.41   & 2.64  & 2.64 & 2.61 & 2.66 & 0.87 & 0.89  &  ...        & ...	  \\
NGC 4030     &17.83 & 16.85 & 18.47  & 17.29   & 3.10  & 3.10 & 3.19 & 3.15 & ...  & ...   & 2.43 (0.06) & 2.66 (0.04) \\
NGC 4258     &15.08 & 14.18 & 15.79  & 14.92   & 3.14  & 3.14 & 3.02 & 2.93 & 0.97 & 0.96  & 29.1 (0.4)  & 24.1 (0.4)  \\
NGC 4303     &16.22 & 15.59 & 17.47  & 16.61   & 2.71  & 2.71 & 3.11 & 3.03 & ...  & ...   & 5.7 (0.9)   & 6.9 (0.1)	 \\
NGC 4536     &17.52 & 16.51 & 18.01  & 17.22   & 4.18  & 4.18 & 4.12 & 4.03 & ...  & ...   & 1.29 (0.02) & 1.86 (0.03) \\
NGC 5005     &16.07 & 14.88 & 16.34  & 15.40   & 3.63  & 3.63 & 3.34 & 3.37 & ...  & ...   & 12.1 (0.2)  & 8.7 (0.4)	 \\
NGC 5054     &17.93 & 16.92 & 18.52  & 17.50   & 3.47  & 3.47 & 3.51 & 3.35 & ...  & ...   & 3.07 (0.08) & 3.35 (0.08) \\
NGC 5055     &14.44 & 13.85 & 15.63  & 15.17   & 3.78  & 3.78 & 3.69 & 3.83 & ...  & ...   & ...	 & ...	   \\
NGC 5248     &17.25 & 16.29 & 17.84  & 17.03   & 3.08  & 3.08 & 3.19 & 3.20 & ...  & ...   & 4.40 (0.08) & 3.8 (0.1)	 \\
NGC 5879     &17.76 & 16.62 & 18.18  & 16.94   & 3.25  & 3.25 & 3.13 & 2.95 & ...  & ...   & 3.0 (0.1)   & 2.26 (0.03) \\
NGC 6384     &18.71 & 17.46 & 19.02  & 17.75   & 3.23  & 3.23 & 3.13 & 3.11 & 0.97 & 0.99  & ...	 & ...	  \\
NGC 6951     &17.64 & 16.71 & 18.30  & 17.35   & 3.54  & 3.54 & 3.64 & 3.60 & 1.07 & 1.12  & 12 (1)	 & 13 (1)	  \\
NGC 7331     &15.07 & 14.19 & 15.83  & 14.97   & 3.17  & 3.26 & 3.26 & 3.20 & ...  & ...   & 1.32 (0.08) & 1.01 (0.02) \\
\enddata
\tablecomments{Magnitudes calculated within circular  apertures. The diameter of the apertures is given in parenthesis. R(a) indicates that the magnitudes are found for annuli of area equivalent to a circular aperture of the given diameter. e.g. `R(a)~50~pc' is the magnitude within an annulus that begins at  25~pc from the nucleus and has an area equivalent to a 50~pc circular aperture. The right-most columns denote the equivalent widths of H$\alpha$ for the sum of the signal within square apertures with sides of 50~pc and 100~pc respectively.}
\label{allsample}
\end{deluxetable}

\begin{figure*}

\includegraphics[width=15.0cm, angle=-90]{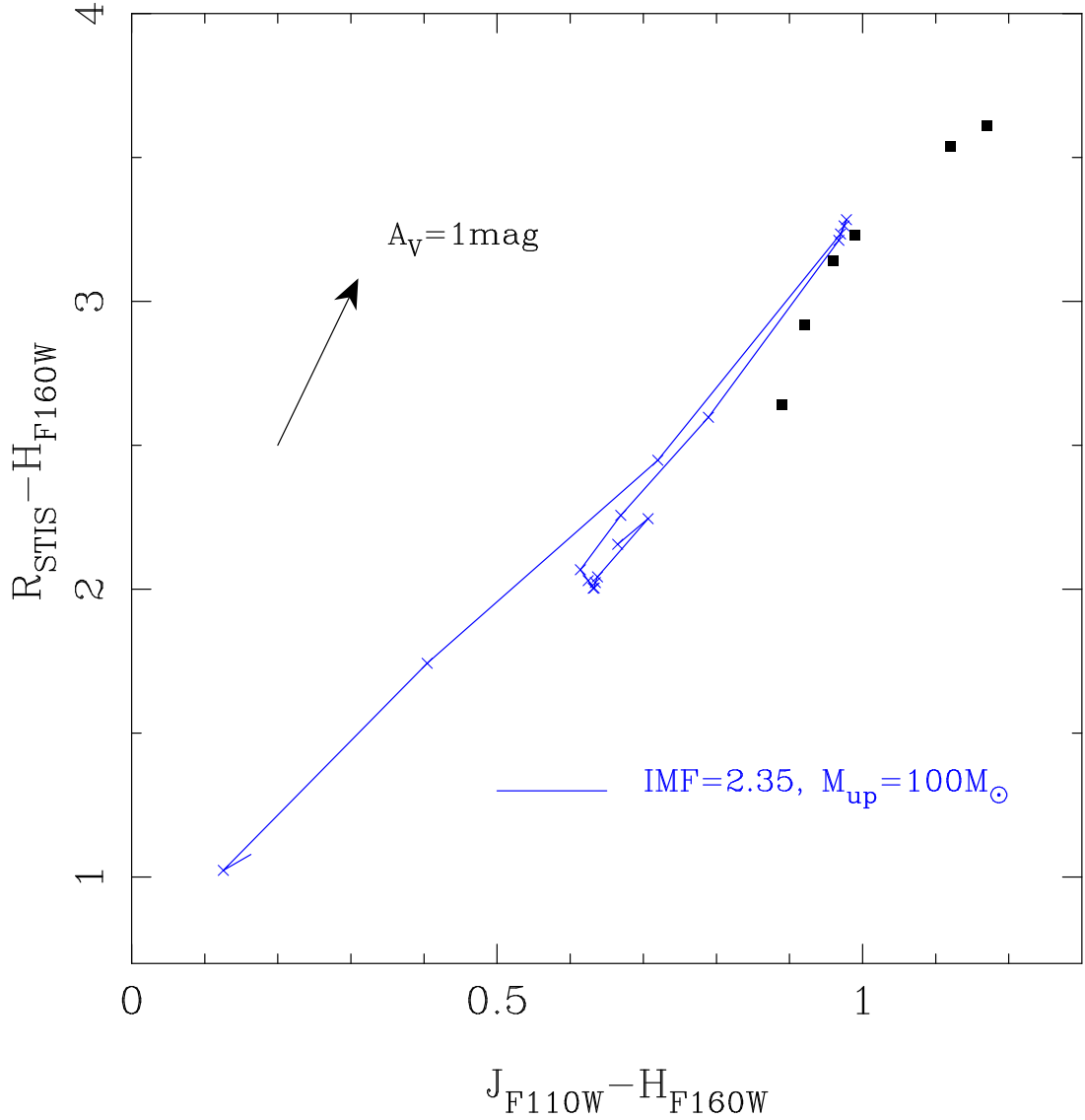}
\caption{The color indices were compared to the models of \citet{leitherer1}. The plot shows the stellar
evolutionary models for an instantaneous burst of star formation, for a Salpeter initial mass function with lower and upper mass cutoffs of 1 and 100\,M$_{\sun}$, respectively.   }
\label{evol1}
\end{figure*}

\begin{figure*}
\epsscale{1}
\plotone{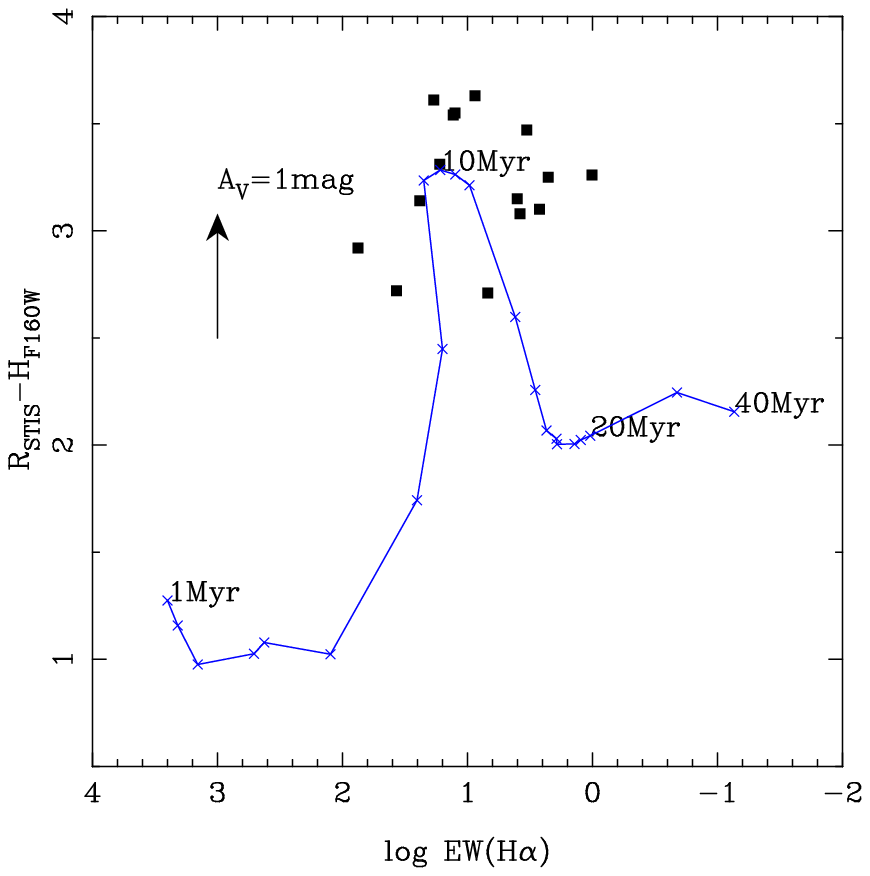}
\caption{The R-H color index against H$\alpha$ equivalent width for 100\,pc apertures placed on the on-nucleus spectra. The model population is the same as in figure ~\ref{evol1}. This is much more useful than the plot in figure ~\ref{evol1} as it suggests the age range of the central stellar populations is $\sim$10-20\,Myr. The points correspond to those galaxies where both color and $H\alpha$ spectral information was available and not just those with obvious central sources apparent in their surface brightness profiles.}
\label{evol2}
\end{figure*}

 As can be seen from Table~\ref{allsample} there is no strong variation of the  color whether the
50\,pc or 100\,pc  aperture is used. Figure~\ref{evol1} shows the R-H against J-H colors for the
100\,pc aperture size.  The time evolution for ages between 1\,Myr (lower left corner) and 20\,Myr (in
intervals of 1\,Myr) and for 30 and 40\,Myr are shown for the three models described above. For
reference the reddest values for the R-H color occur for ages of between 9 and  12\,Myr, which
correspond to the epoch when the first supergiants appear and start contributing significantly to the
light in the H-band. We also show in this diagram the effect of one magnitude of  extinction. 
Unfortunately the results cannot distinguish between the models.  The reddening vector is almost
parallel to the models and so with the right amount of extinction the points could fit a variety of
ages.  Clearly the R-J and J-H colors alone cannot be used to constrain the age of the  stellar
population.

A better way to constrain the age of the stellar population is to use the equivalent 
width of H$\alpha$. Table~\ref{allsample} shows the EW within square apertures defined  by sides of  50\,pc (or 100\,pc) and the slit width of $0.2\arcsec$. As can be seen from figure~83 in \citet{leitherer1}, 
for instantaneous star formation the EW of H$\alpha$ decreases with time.   
Figure~\ref{evol2} shows the R-H color against H$\alpha$ equivalent width. Here we plot
the data for the 14 galaxies with measured colors and EWs. Again the evolution is shown
for ages between 1 and 20\,Myr (in 1\,Myr intervals) and 30 and 40\,Myr.
Note that in this case the extinction vector only affects the color because we are assuming 
that the extinction to the stars and the gas is the same, and thus the equivalent width of 
the line is independent of the extinction. The H$\alpha$ equivalent width allows the ages of the stellar 
population to be constrained. We can also infer the extinction to the central 
100\,pc. The typical values for our sample are $A_V=1-2\,$\,mag. The derived ages of the central 
sources (they are mostly independent of the model assumed) are typically between 10 and 15\,Myr, except
for NGC~3310  where the age is 6-7\,Myr.

The case of continuous star formation was ruled out since the Starburst99 \citep{leitherer1} simulations indicate that the equivalent width of H$\alpha$ should remain greater than 146\,\AA, which is not the case for any of the galaxies in our sample. We also believe that the case of an older stellar population plus  continuous star formation is not the case in the central regions due to the similarity in colors in both 50\,pc and 100\,pc apertures.

\begin{deluxetable}{lccccc} 
\tabletypesize{} 
\tablewidth{0pt}
\tablecaption{Line Flux ratios relative to H{$\alpha$}}
\tablehead{Galaxy &  Activity   &[\ion{N}{2}]      &[\ion{N}{2}]	  &[\ion{S}{2}]    & [\ion{S}{2}]   \\   
	    Name  &  Status   & (6549.85\,\AA)   &  (6585.28\,\AA)   & (6718.29\,\AA) &  (6731.67\,\AA)} 
		  
\startdata 
NGC1300 & ... (...)                   &  0.83  (0.03)  &  1.83  (0.03)  & 0.82  (0.02)   &  0.65  (0.02)   \\
NGC2748 & ... (\ion{H}{2})            &  0.182 (0.003) & 0.370 (0.005)  & ...	         &   ...           \\
NGC2903 & \ion{H}{2} (\ion{H}{2})     &  0.16  (0.01)  & 0.507 (0.004)  & ...	         &   ...	   \\
NGC2964 & \ion{H}{2} (\ion{H}{2})     &  0.211 (0.002) & 0.556 (0.002)  & 0.184 (0.003)  &  0.174 (0.003)  \\
NGC3310 & \ion{H}{2} (\ion{H}{2})     &  0.192 (0.002) & 0.587 (0.002)  & 0.121 (0.001)  &  0.125 (0.002)  \\
NGC4258	& \ion{H}{2} (\ion{H}{2})     &  0.49  (0.02)  & 0.77  (0.01)   & 0.214 (0.003)  &  0.24  (0.003)  \\
NGC4303	& \ion{H}{2}/S2 (\ion{H}{2})  &  1.886 (0.006) & 1.044 (0.004)  & 0.487 (0.002)  &  0.456 (0.004)  \\
NGC4536	& \ion{H}{2} (\ion{H}{2})     &  ...	       & 1.81  (0.02)   & 0.53  (0.01)   &  1.06  (0.02)   \\
NGC5005 & S2/L (L1.9)                 &  2.17  (0.07)  & 5.0   (0.2)    & 1.10  (0.04)   &  1.57  (0.05)   \\
NGC6951 & L/S2 (S2)                   &  1.36  (0.02)  & 4.30  (0.05)   & 0.69  (0.01)   &  0.79  (0.01)   \\
\enddata		   
\tablecomments{Line Flux ratios, calculated by dividing the corresponding  H{$\alpha$} (6564.61\AA) flux. All wavelengths are for a vacuum. Uncertainties were calculated by repeating each flux measurement a minimum of 10 times and are shown in parenthesis. The current activity status values are repeated from Table~\ref{little_sample} and are taken from the NASA Extragalactic Database (NED) with classification shown in
parenthesis taken from \citet{ho1}, where \ion{H}{2}=\ion{H}{2} nuclei, S=Seyfert nuclei and L=LINER. }
\label{line_ratios}
\end{deluxetable}

Note however that the underlying assumption is that all the ionizing photons come from stars.  If an active galactic nucleus were present then this
would tend to increase the equivalent width and give the impression that a stellar population was actually younger than its real age. We cannot
unambiguously rule out previously unknown AGN in the galaxies in our sample. As a check, using our data we examined the line ratios of both
[N~II]~($\lambda$6550,6585~\AA) and [S~II]~($\lambda$6718,6732~\AA) to that of H~$\alpha$~($\lambda$6565~\AA). The results are presented in
Table~\ref{line_ratios}. Two possible additional weak AGN, indicated by the relatively high values of the [\ion{N}{2}] ratios are NGC~1300 and
NGC~4536. It should also be noted that the galaxies with known activity in our sample have EWs that are well spread throughout the distribution
of equivalent widths (e.g. EW=24.1$\pm$0.4  for NGC~4258 and EW=1.01$\pm$0.02 for NGC~7331).  In summary, while we think it is unlikely that the
equivalent widths are dominated by ionization from weak nuclear activity we cannot rule it out with this dataset.

\section{Summary}

Using NICMOS  images we constructed surface brightness profiles, to which we fitted Nuker laws to search for the presence of nuclear star
clusters. Several of the images had previously been fit by \citet{seigar1} and there is some variation in their resultant Nuker parameters when
compared with ours partly due to the necessary subjectivity in performing the fits. In 4 cases; NGC~2748, NGC~3949, NGC~4030 and NGC~4303  there
appears to be an excess which can not be accounted for by the Nuker profile alone.  In the case of NGC~2748 the apparent excess is misleading and is probably due to obscuring dust causing an inflexion in the surface brightness profile. However, for the latter 3 galaxies the excesses are likely to be real and may be due to the presence of nuclear star clusters. 

Although star clusters and AGN can coexist (e.g. \citealt{colina1}) we attempted to identify any  unknown AGN by using [\ion{N}{2}]  and
[\ion{S}{2}]  line flux ratios (relative to  H{$\alpha$}) and found tentative evidence for weak AGN in NGC~1300 and NGC~4536.

As a first step towards constraining the ages of the central stellar populations we used color information from the STIS  and NICMOS images and the equivalent width of H$\alpha$ for
the STIS   spectra and compared these data to the stellar evolutionary models of \citet{leitherer1}. The equivalent width of H$\alpha$ was the
important factor as it constrained the ages to $\sim$10\,Myr provided that the ionization of the disk was primarily due to the central stellar
population. To confirm the age of the population we would need to perform spectroscopy on the galaxy centers.

\section{Acknowledgments}  MAH thanks A Robinson and R. van der Marel for their comments and suggestions. MAH  particularly thanks Jim Collett for
very useful discussions. We would also like to thank the anonymous referee for valuable suggestions which improved this paper. We have made
use of the LEDA database (http://leda.univ-lyon1.fr). This research has made use of the NASA/IPAC Extragalactic Database (NED) which is operated
by the Jet Propulsion Laboratory, California Institute of Technology, under contract with the National Aeronautics and Space Administration.

\end{document}